\documentclass[10pt,twoside]{Lowx2011}
\usepackage{epsf,amsmath}
\usepackage{graphicx}

\newcommand{\bce}{\begin{center}} 
\newcommand{\ece}{\end{center}}
\newcommand{\beq}{\begin{equation}}
\newcommand{\eeq}{\end{equation}}
\newcommand{\bea}{\vspace{0.25cm}\begin{eqnarray}}
\newcommand{\eea}{\end{eqnarray}}

\newcommand{\brho}{\mbox{\boldmath $\rho$}}

\newcommand{\bk}{{\bf k}}

\newcommand{\bq}{{\bf q}}
\newcommand{\br}{{\bf r}}

\newcommand{\ba}{\begin{array}}
\newcommand{\ea}{\end{array}}

\def\lsim{\mathrel{\rlap{\lower4pt\hbox{\hskip1pt$\sim$}}
    \raise1pt\hbox{$<$}}}         
\def\gsim{\mathrel{\rlap{\lower4pt\hbox{\hskip1pt$\sim$}}
    \raise1pt\hbox{$>$}}}         

\def\Pom{{\bf I\!P}}

\def\beq{\begin{equation}}
\def\endeq{\end{equation}}
\def\arr{\begin{eqnarray}}
\def\endarr{\end{eqnarray}}

\begin{document}

\title{ UHE NEUTRINOS: FUSING  GLUONS  WITHIN  DIFFRACTION CONE}

\author{R.~FIORE and V.R.~ZOLLER\\ \\
 \\ \\
Institute for  Theoretical and Experimental Physics,
Moscow 117218, Russia\\
E-mail: zoller@itep.ru\\ \\
Dipartimento di Fisica,
Universit\`a     della Calabria\\
and\\
 Istituto Nazionale
di Fisica Nucleare, Gruppo collegato di Cosenza,\\
I-87036 Rende, Cosenza, Italy\\
E-mail: fiore@cs.infn.it}

\maketitle

\begin{abstract}
Currently available estimates of the gluon-fusion effect in ultra-high 
energy neutrino-nucleon interactions as well as  in DIS on protons
  suffer from   uncertainty
in defining  the scattering profile function $\Gamma(b)$. Indeed, 
the area, $S$, in the impact parameter space  populated with 
interacting gluons  varies  by a factor of  $4 - 5$ from one analysis to 
another.  To get rid of uncertainties we specify  the  dipole-nucleon 
partial-wave amplitude $\Gamma(b)$ which  meets the restrictions  
imposed by both  the total dipole-nucleon  cross section and the 
small angle elastic scattering amplitude. The area $S$ becomes a
 well defined quantity proportional to  the diffraction cone slope.
We solve numerically the non-linear color dipole BFKL equation
and  evaluate  the UHE neutrino-nucleon total cross section.
Our finding  is that the saturation is  a rather weak effect, $\lsim 25\%$,
 up to $E_{\nu}\sim 10^{12}$ GeV.

\end{abstract}



\markboth{\large \sl \hspace*{0.25cm}\underline{V.R.~ZOLLER} \& R.~FIORE
\hspace*{0.25cm} Low-$x$ Meeting 2011} 
{\large \sl \hspace*{0.25cm} THE
LOW$x$ 2011 MEETING PROCEEDINGS}

\section{Introduction}
Practical needs of the neutrino astrophysics \cite{NeuAs}
inspired many papers on the cross section $\sigma^{\nu N}(E_{\nu})$  
for the scattering of 
Ultra-High Energy (UHE) neutrinos on nucleons and nuclei.
The question of interest is the interplay of unitarity constraints on 
$\sigma^{\nu N}(E_{\nu})$ and the evolution of QCD  parton densities.
The  UHE neutrinos 
probe the gluon density in the target nucleon   at very small 
values of 
Bjorken $x$.  The BFKL \cite{BFKL} distribution of gluons 
 grows fast to smaller 
$x$, $G(x,Q^2)\propto x^{-\Delta}$, where, phenomenologically, 
  $\Delta\approx 0.3$. 
Hence, the neutrino-nucleon cross sections $\sigma_{\nu N}\sim E_{\nu}^{\Delta}$ violating  the Froissart bound.

 The original idea of 
Ref.\cite{Kancheli73} developed further in Ref.\cite{NZ75} 
was that 
an  overlap in transverse space  and recombination of partons leads 
to  a slow down of the growth of the parton density and finally  to  
the  saturation of parton densities. 
Quantitative QCD analysis of the non-linear  effects in terms of 
the  gluon density $G(x,Q^2)$ was initiated by 
Ref.  \cite{GLRMQ}.
More recently,  different derivations of the non-linear BFKL equation were
 presented in \cite{B}.  
The strength of the saturation effect  is usually  estimated  as 
\beq
\kappa\sim{\sigma(x,Q^2)/S},
\label{eq:SATPAR}
\eeq
where  $\sigma$ stands for some gluon interaction cross section.
The radius of 
the area in the impact parameter plane, $S=\pi R^2$, within 
which the interacting gluons are expected to be distributed,
varies considerably, from $R^2=16$ GeV$^{-2}$ in \cite{KK2003} down to
$R^2=3.1 $ GeV$^{-2}$ in \cite{Bartels}.
Besides, the area $S$ is assumed to be independent of $x$.  However,
 because of the BFKL diffusion property,
under certain conditions,  $S$ 
acquires  the Regge 
contribution $\sim \alpha^{\prime}_{\Pom}\log(1/x)$ \cite{NZZ97}.

In this communication we
specify  the  dipole-nucleon partial-wave amplitude 
  which  meets the restrictions  imposed by 
both  the total dipole-nucleon  cross section and the 
small angle elastic scattering amplitude. The area $S$ becomes a  
well defined
quantity proportional to the diffraction cone slope.
We solve numerically the non-linear color dipole BFKL equation
and  evaluate  the UHE neutrino-nucleon total cross section.
 Our finding  is that 
 the saturation is  a rather weak effect, $\lsim 25\%$,
 up to $E_{\nu}\sim 10^{12}$ GeV.
This result differs
  from  predictions found  in extensive  literature  
on the subject \cite{KK2003,NeuSec}.

\section{The partial-wave amplitude and diffraction cone slope}

Generalization of the color dipole BFKL approach developed 
in \cite{NZZJL94}  
to the equation for diffraction slope
$B(x,r)$ proceeds as follows \cite{SL94,SLPL}.
In the impact-parameter representation the imaginary part of the
 elastic dipole-nucleon amplitude reads
\beq
{\cal A}(\xi,r,\bq)=2\int d^2{\bf b}\exp(-i\bq\bk) \Gamma(\xi,r,{\bf b})
\label{eq:AMPL}
\eeq
and the dipole cross section is
$
\sigma(\xi,r)=2\int d^2{\bf b}\, \Gamma(\xi,r,{\bf b}).
$
The diffraction slope 
$
B=-2(d\log {\cal A}/dq^2)|_{q^2=0}
$
 for the forward cone is
\beq
B(\xi,r)= {1\over 2}\langle {\bf b}^2\rangle =
{1\over\sigma(\xi,r)}\int d^2{\bf b}~  {\bf b}\,^2~\Gamma(\xi,r,{\bf b}),
\label{eq:BASAV}
\eeq
where
$\Gamma(\xi,r,{\bf b})$ is the profile function and
${\bf b}$ is the impact parameter defined with respect
to the center of the $q$-$\bar{q}$ dipole. In the $q\bar{q}g$
state, the $qg$ and $\bar{q}g$ dipoles
have the impact parameter ${\bf b}+{\brho}_{2,1}/2$.
Then \cite{SL94}, 
\bea
{\partial \Gamma(\xi,r,{\bf b})\over \partial \xi}
={N_c^2\over N_c^2-1} \int d^{2}{\brho}_{1}
\left|\psi({\brho}_{1})-\psi({\brho}_{2})\right|^{2}\nonumber\\
\times\left[\Gamma(\xi,\rho_{1},{\bf b}+{1\over 2}{\brho}_{2}) +
\Gamma(\xi,\rho_{2},{\bf b}+{1\over 2}{\brho}_{1}) -
\Gamma(\xi,r,{\bf b})\right]
\label{eq:GAMBFKL}
\eea
where $\xi=\log(x_0/x)$ and $\psi({\brho})$ is the radial light 
cone wave function of the $qg$
 dipole with the Yukawa screening of infrared gluons \cite{NZZJL94}
\beq
\psi({\brho})={\sqrt{C_F\alpha_S(R_i)}\over \pi}{{\brho}\over \rho R_c}
K_{1}(\rho/R_c).
\label{eq:PSIQG}
\eeq
Here $\brho_{1,2}$ are the $q$-$g$ and $\bar{q}$-$g$ separations
in the two-dimensional impact parameter plane
for dipoles generated by the $\bar{q}$-$q$
color dipole source,
${\bf r}$ is the $\bar{q}$-$q$ separation
and
$K_{\nu}(x)$ is the modified Bessel function. At 
$r,\rho_{1,2} \ll R_{c}$ and
in the $\alpha_{S}=const$ approximation, the scaling BFKL equation 
\cite{BFKL} is
obtained.

The BFKL dipole cross section $\sigma(\xi,r)$
sums the Leading-Log$(1/x)$ multi-gluon production cross
sections. Consequently,  a
realistic boundary condition for the
BFKL dynamics is  the two-gluon exchange Born amplitude at $x=x_0=0.03$.
The running QCD coupling $\alpha_{S}(R_i)$ must be taken at 
the shortest relevant
distance $R_{i}={\rm min}\{r,\rho_{i}\}$ and
in the numerical analysis  an infrared freezing 
$\alpha_{S}(q^{2}) \leq 0.8$
has been imposed. In Ref.\cite{Lipatov86} it was found that incorporation  of 
the asymptotic freedom into the BFKL equation  splits the 
 cut in the complex $j$-plane into a series of 
 isolated BFKL-Regge  poles. 
Also, in \cite{NZZ97} it was shown that
breaking of scale invariance by a running
$\alpha_{S}(r)$ supplemented by
the finite gluon propagation radius $R_{c}$,  changes
the  nature of the BFKL pomeron from a {\sl fixed} cut
 to a series of {\sl moving}
poles with the finite Regge slope $\alpha^{\prime}_{\Pom}$ of the pomeron
trajectory $j=\alpha(t)=\alpha_{\Pom}(0)+\alpha^{\prime}_{\Pom}t.$
 The
preferred choice $R_{c}=0.27$\,fm  gives 
$\alpha^{\prime}_{\Pom}=0.072$ GeV$^{-2}$,
$\Delta_{\Pom}=\alpha_{\Pom}(0)-1=0.4$  and leads to a very good
description  of the  data on the proton structure 
function and the diffraction cone slope \cite{CharmBeauty}
 at small $x$.

In \cite{SLPL} the diffraction slope $B$ 
in the elastic scattering of the  dipole $r$ on the nucleon
(with the gluon-probed radius $R_N$)
was presented in a very symmetric form  
\beq 
B(\xi,r)={1\over 8}r^2+{1\over 3} R_N^2 
+2\alpha^{\prime}_{\Pom}\xi.
\label{eq:BSLOPE}
\eeq
\begin{figure}[hbt]
\centering
\vskip-0.25cm
\includegraphics[width=.50\hsize]{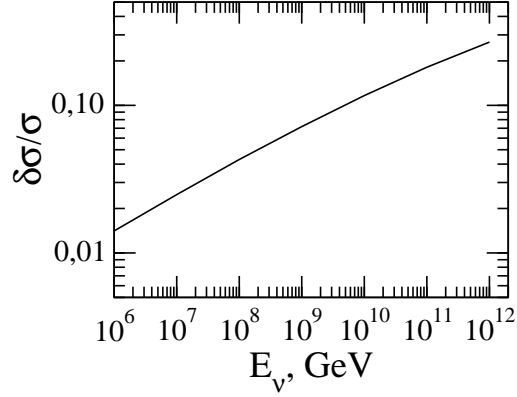}
\vskip -0.3cm
\caption{The non-linear correction to the 
neutrino-nucleon cross section 
 as a function 
of neutrino energy}
\end{figure}

\section{Non-linear effects}
Following \cite{B}, one can add to the $rhs$ of Eq.~(\ref{eq:GAMBFKL}) 
(assuming $N_c\gg 1$) the term
\beq
-\Gamma(\xi,\rho_{2},{\bf b}+{1\over 2}{\brho}_{1})
\Gamma(\xi,\rho_{1},{\bf b}+{1\over 2}{\brho}_{2}).
\label{eq:BK}
\eeq
The  partial-wave amplitude in the impact parameter space  reads
\beq
\Gamma(\xi,\rho,{\bf b})={\sigma(\xi,r)\over 4\pi B(\xi,r)}
\exp\left[-{b^2\over 2B(\xi,r)}\right].
\label{eq:GAMMA}
\eeq
Then, integrating over ${\bf b}$ in Eqs.(\ref{eq:GAMBFKL},\ref{eq:BK}) 
yields
\bea
{\partial \sigma(\xi,r) \over \partial \xi} =
 \int d^{2}{\brho}_{1}\,\,
\left|\psi({\brho}_{1})-\psi({\brho}_{2})\right|^{2}\nonumber\\
\times\left\{\sigma(\xi,\rho_{1})+
\sigma(\xi,\rho_{2})-\sigma(\xi,r)\right.\nonumber\\
\left.-{\sigma(\xi,\rho_{1})\sigma(\xi,\rho_{2})\over 4\pi(B_1+B_2)}
\exp\left[-{\br^2\over 8(B_1+B_2)}\right]\right\}.
\label{eq:BFKLNL}
\eea
Shown in Fig.~1  is  the ratio
$
\delta\sigma/\sigma = 1-\sigma^{nl}_{\nu N}/\sigma^{l}_{\nu N}
$
which quantifies the strength of non-linear effects in the $\nu N$ cross 
section.
Here $\sigma^{nl}_{\nu N}$ represents the observable $\nu N$ charged current 
cross section
obtained from the numerical solution of Eq.(\ref{eq:BFKLNL}). 
The cross section
$\sigma^{l}_{\nu N}$ derives from the solution of Eq.~(\ref{eq:BFKLNL})
 with the non-linear term turned off .
The ${\nu N}$ cross section calculated for $1/R_c=0.75$ GeV and
$R_N^2=12$ GeV$^{-2}$ is shown in  Fig.~2, where $\sigma_{\nu N}$
is identical to $\sigma^{nl}_{\nu N}$.
We conclude, that the non-linear effect is rather weak, $
\delta\sigma/\sigma\approx 0.25$ at $E_{\nu}=10^{12}$ GeV. 
This result differs considerably from earlier predictions.

\begin{figure}[h]
\centering
\includegraphics[width=.55\hsize]{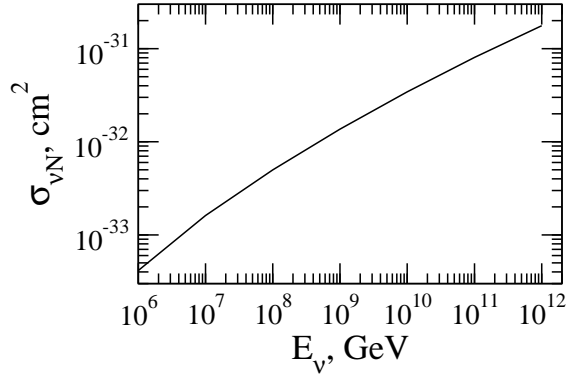}
\vskip -0.3cm
\caption{The charged current neutrino-nucleon cross section
as a function of neutrino energy}
\end{figure}

\section*{Acknowledgments} We thank all organizers of 
Low-$x$ Meeting 2011 for their warm hospitality.  
We are indebted to N.N. Nikolaev for useful comments.
This work was
supported in part by the RFBR grants 09-02-00732, 11-02-00441
and the DFG grant 436 RUS 113/991/0-1.

\end{document}